\documentclass[aps,showpacs,twocolumn]{revtex4}
\usepackage{epsfig}
\usepackage{graphicx,color,dcolumn}

\begin{document}
\author{You-chang Yang$^{1,2}$, Zurong Xia$^1$, Jialun Ping$^1$}
\email{jlping@njnu.edu.cn}
\affiliation{$^1$Department of
Physics, Nanjing Normal University, Nanjing 210097, P. R.
China\\$^2$Department of Physics, Zunyi Normal College, Zunyi 563002, P.
R. China}

\title{Are the $X$(4160) and $X$(3915) charmonium states?}

\begin{abstract}
Inspired by the newly observed $X$(4160) and $X$(3915) states,
we analyze the mass spectrum of these states in different quark
models and calculate their strong decay widths by the $^3P_0$ model.
According to the mass spectrum of charmonium states predicted by the
potential model, the states
$\chi_0(3^3P_0),~\chi_1(3^3P_1),~\eta_{c2}(2^1D_2),~\eta_c(4^1S_0)$
all can be candidates for the $X$(4160). However, only the decay width
of the state $\eta_{c2}(2^1D_2)$ in our calculation is in good agreement
with the data reported by Belle and the decay of
$\eta_{c2}(2^1D_2)\to D\bar{D}$, which is not seen
in experiment, is also forbidden. Therefore, it is reasonable to interpret the
charmonium state $\eta_{c2}(2^1D_2)$ as the state $X(4160)$.
For the state $X(3915)$, although the mass of $\chi_0(2^3P_0)$ is
compatible with the experimental value, the calculated strong decay width is much
larger than experimental data. Hence, the assignment of $X(3915)$ to
charmonium state $\chi_0(2^3P_0)$ is disfavored in our calculation.
\end{abstract}

\maketitle
\section{Introduction} \label{introduction}
Many new charmonium like states, the so-called $XYZ$ mesons, have
been reported  by Belle and BaBar collaborations in recent years.
Some of these states can be understood as conventional mesons that
are comprised of only pure $c\bar{c}$ quark pair. However, most of the
$XYZ$ states do not match well the mass spectrum of
$c\bar{c}$ predicted by the QCD-motivated potential models. By
considering the effects of virtual mesons loop
\cite{Eichtenprd17,Eichtenprd73,Barnesprc77,chaoprd80} and color
screening \cite{prd79094004}, the masses of some excited charmonium
states are smaller than it calculated by conventional quark model.
Therefore, some $XYZ$ states \cite{Eichtenprd73} may be still compatible with the
mass spectrum of charmonium. However, the state $X(3872)$
\cite{Eichtenprd73,chaoprd80,Suzuki72} is probably the most robust of
all the charmonium like objects.

Last year, Belle collaborations reported a new charmonium like
state, the $X(4160)$ \cite{X4160}, in the processes $e^+e^- \to
J/\psi D^{(*)}\bar{D}^{(*)}$ with a significance of 5.1$\sigma$. It
has the mass $M=4156^{+25}_{-20}\pm 15$ MeV, and width
$\Gamma=139^{+111}_{-61}\pm 21$ MeV.  Based on the the processes
$e^+e^- \to J/\psi D\bar{D}$, $e^+e^- \to J/\psi D^{*}\bar{D}$, and
$e^+e^- \to J/\psi D^{*}\bar{D}^{*}$, The upper
limits of the branch ratios of $X(4160)$ are given as,
\begin{eqnarray*}
&& \mathcal{B}_{D\bar{D}}(X(4160))/\mathcal{B}_{D^*\bar{D}^*}(X(4160))<
0.09 , \\
&& \mathcal{B}_{D^*\bar{D}}(X(4160))/\mathcal{B}_{D^*\bar{D}^*}(X(4160))<
0.22.
\end{eqnarray*}
The $X(4160)$ has possible charge parity $C=+$ mostly, since the
photon $\gamma$ and $J/\psi$ have $J^{PC}=1^{--}$, and $e^+e^- \to
\gamma \to J/\psi X(4160)$ is a main process. Hence the $X(4160)$
can have
$J^{PC}=0^{-+}$, $0^{++}$, $1^{-+}$, $2^{-+}$, $1^{++},~2^{++},\ldots$. In
Ref.\cite{chaox4160}, Chao discussed the possible interpretation of
 the $X(4160)$ in view of production rate in $e^+e^- \to J/\psi
X(4160)$. He believes that the charmonium states $4^1S_0,~3^3P_0$
may be assigned to the state $X(4160)$ by analogy with the cross
section of $e^+e^- \to J/\psi\eta_c(1S) (\eta_c(2S) \chi_{c0}(1P))$,
while the $2^1D_2$ \cite{chao77014002} can not be rule out.
According to the mass spectrum \cite{prd79094004} predicted by the
potential model with color screening, Li and Chao also give some
arguments about the $\chi_0(3^3P_0)$ as an assignment for the
$X(4160)$.

Using the vector-vector interaction within the framework of the
hidden gauge formalism, Molina and Oset \cite{Oset0907} suggested that
the $X(4160)$ is a molecular state of $D_s^*\bar{D}_s^*$ with
$J^{PC}=2^{++}$.

Very recently, Refs.\cite{Olsen0909,CZYuan0910,Zupanc0910,Gordfrey0910}
reported the newest charmonium like state, the $X(3915)$, which is observed by
Belle in $\gamma\gamma \to \omega J/\psi$ with a statistical
significance of 7.5$\sigma$. It has the mass and width
\[
M=3914\pm4\pm2~\text{MeV},~~~ \Gamma=28\pm 12^{+2}_{-8}~\text{MeV}.
\]
Belle collaborations determine the $X(3915)$ production rate
$\Gamma_{\gamma\gamma}(X(3915))$ $\mathcal{B}(X(3915)\to \omega
J/\psi)=69\pm16^{+7}_{-18}$ eV and
$\Gamma_{\gamma\gamma}(X(3915))$ $\mathcal{B}(X(3915)\to \omega
J/\psi)=21\pm 4^{+2}_{-5}$ eV for $J^P=0^+$ or $2^+$, respectively.
Because the partial width of this state to $\gamma\gamma$ or $\omega
J/\psi$ is too large, it is very unlikely to be a charmonium state
analyzed by Yuan \cite{CZYuan0910}.

The $X(3915)$ also has the charge parity $C=+$, because it is
observed in the process of $\gamma\gamma  \to  \omega J/\psi$. In
Ref.\cite{XiangLiu}, Liu \textit{et al.} argued that the $\chi_0(2^3P_0)$
can be assigned to the $X(3915)$ if taking $R=1.8 \sim 1.85$ GeV$^{-1}$
in the SHO (the simple harmonic oscillator wave functions).

Up to now, the interpretation of the $X(4160)$ and $X(3915)$ is
still unclear. The states
$\chi_0(3^3P_0)$, $\chi_1(3^3P_1)$, $\eta_{c2}(2^1D_2)$ listed in
Table \ref{ccspectrum} all can be interpreted as the $X(4160)$ just on
mass level. Which charmonium state is an assignment for the
$X(4160)$? One can answer this question in different ways. We study
the $X(4160)$ and $X(3915)$ via strong decay by the $^3P_0$ model
\cite{NPB10521,Yaouanc,PRD546811,Roberts1} in this work. In
following discussion, we take the
$\chi_0(3^3P_0),~\chi_1(3^3P_1),~\eta_{c2}(2^1D_2),~\eta_c(4^1S_0)$
and $\chi_0(2^3P_0)$ as candidates of the $X(4160)$ and $X(3915)$,
respectively.
\begin{ruledtabular}
\begin{table}[htb]
\caption{Theoretical mass spectrum of the charmonium candidates for
the $X(4160)$ and $X(3915)$. The mass are in units of MeV. The
results are taken from Ref.\cite{prd79094004} with color screening
potential model, and Ref.\cite{prd72054026} including 
Nonrelativistic potential and Godfrey-Isgur relativized potential model. \label{ccspectrum}}
\begin{tabular}{cccccc}
 State & $\chi_0(2^3P_0)$ & $\eta_c(4^1S_0)$ & $\chi_0(3^3P_0)$ &
  $\chi_1(3^3P_1)$ & $\eta_{c2}(2^1D_2)$ \\
 $J^{PC}$ & $0^{++}$ &  $0^{-+}$ & $0^{++}$  &
  $1^{++}$ &  $2^{-+}$ \\
\hline
Ref.\cite{prd79094004} SCR  & 3842  & 4250  & 4131  & 4178  & 4099  \\ \hline
Ref.\cite{prd72054026} NR & 3852 & 4384 & 4202 & 4271 & 4158 \\
Ref.\cite{prd72054026} GI~ & 3916 & 4425 &  4292 & 4317 & 4208 \\
\end{tabular}
\end{table}
\end{ruledtabular}

The paper is organized as follows. In the next section we take a
review of the $^3P_0$ model. Sect. \ref{secamp} devotes to
discuss the possible strong decay channels and gives the
corresponding amplitudes of the candidates for the $X(4160)$ and
$X(3915)$. In Sect. \ref{discussion} we present and analyze the
results obtained by the $^3P_0$ model. Finally, the summary of the
present work is given in the last section.

\section{A review of the $^3P_0$  model of meson decay \label{3p0model}}

\begin{center}
\epsfig{file=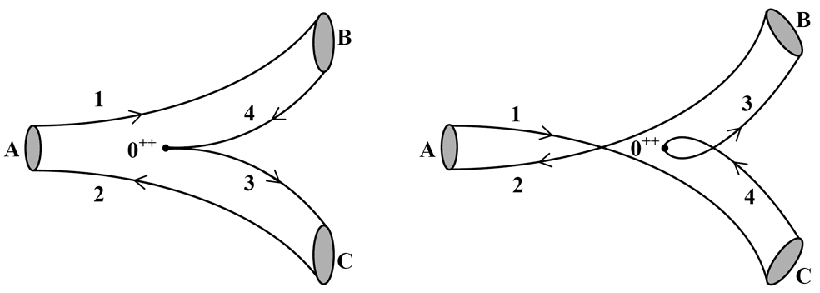,width=7.0cm}

{\small Fig.1 The two possible diagrams contributing to $A\rightarrow B+C$ in the $^3P_0$
 model.}
\end{center}

The $^3P_0$ decay model, also known as the Quark-Pair Creation model
(QPC), was originally introduced by Micu\cite{NPB10521} and further
developed by Le Yaouanc, Ackleh, Roberts \textit{et
al}.\cite{Yaouanc,PRD546811,Roberts1}. It is applicable to OZI
(Okubo, Zweig and Iizuka) rule allowed strong decays of a hadron into
two other hadrons, which are expected to be the dominant decay modes
of a hadron. Due to the $^3P_0$ model gives a good description of
many observed partial widths of the hadrons, it has been widely used
to evaluate the strong decays of mesons and baryons composed of
$u,~d ,~s,~c,~b$ quarks
\cite{Close,XiangLiu,Swanson,zhusl,XiangLiu2,prd342809,prd506855,Blundell,prd35907,prd554157,prd68054014,prd62014011,ldm}.
The $^3P_0$ model of strong decays assumes that quark-antiquark pair
are created with vacuum quantum number $J^{PC}=0^{++}
$\cite{NPB10521}. The diagrams of all possible decay process
$A\rightarrow B+C$ of meson are shown in Fig.1. In
many cases only one of them contributes to the strong decay of
meson.

The transition operator of this model takes
\begin{eqnarray}
&&T=-3~\gamma\sum_m\langle 1m1-m|00\rangle\int
d\mathbf{p}_3d\mathbf{p}_4\delta^3(\mathbf{p}_3+\mathbf{p}_4)\nonumber\\
&&~~~~\times{\cal{Y}}^m_1(\frac{\mathbf{p}_3-\mathbf{p}_4}{2})
\chi^{34}_{1-m}\phi^{34}_0\omega^{34}_0b^\dagger_3(\mathbf{p}_3)d^\dagger_4(\mathbf{p}_4),
\end{eqnarray}
where $\gamma$, which is a dimensionless parameter, represents the
probability of the quark-antiquark pair created from the vacuum and
can be extracted by fitting observed experimental data.
$\phi^{34}_{0}=(u\bar{u}+d\bar{d}+s\bar{s})/\sqrt{3}$, ~
$\omega^{34}_0=(R\bar{R}+G\bar{G}+B\bar{B})/\sqrt{3}$ are flavor and
color singlet state, respectively. $\chi_{{1,-m}}^{34}$ is a
spin-triplet state.
${\cal{Y}}^m_l(\mathbf{p})\equiv|p|^lY^m_l(\theta_p,\phi_p)$ is the
$l$th solid harmonic polynomial that reflects the momentum-space
distribution of the created quark-antiquark pair.
$b^\dagger_3(\mathbf{p}_3)$, $d^\dagger_4(\mathbf{p}_4)$ are the
creation operators of the quark and antiquark, respectively.

In general, the mock state is adopted to describe the meson with the
spatial wave function
$\psi_{n_AL_AM_{L_A}}(\mathbf{p}_1,\mathbf{p}_2)$ in the momentum
representation \cite{mock}.
\begin{widetext}
\begin{eqnarray}
|A(n_A{}^{2S_A+1}L_{A}\,\mbox{}_{J_A M_{J_A}})(\mathbf{P}_A)\rangle
&\equiv& \sqrt{2E_A}\sum_{M_{L_A},M_{S_A}}\langle L_A M_{L_A} S_A
M_{S_A}|J_A
M_{J_A}\rangle\nonumber\\
&&\times  \int
d\mathbf{p}_A\psi_{n_AL_AM_{L_A}}(\mathbf{p}_1,\mathbf{p}_2)\chi^{12}_{S_AM_{S_A}}\phi^{12}_A\omega^{12}_A|q_1(\mathbf{p}_1)\bar{q}_2(\mathbf{p}_2)\rangle,
\end{eqnarray}
with the normalization conditions
\begin{eqnarray}
\langle A(n_A{}^{2S_A+1}L_{A}\,\mbox{}_{J_A M_{J_A}})(\mathbf{P}_A)
|A(n_A{}^{2S_A+1}L_{A}\,\mbox{}_{J_A
M_{J_A}})(\mathbf{P^\prime}_A)\rangle=2E_A\delta^3(\mathbf{P}_A-\mathbf{P^\prime}_A).
\end{eqnarray}
\end{widetext} where $n_A$ represent the radial quantum number of the
meson $A$ composed of $q_1,~\bar{q}_2$ with momentum $\mathbf{p_1}$
and $\mathbf{p_2}$. $E_A$ is the total energy, $\mathbf{P}_A$ is
the momentum of the meson $A$ and $\mathbf{p}_A=(m_1\mathbf{p}_1-m_1\mathbf{p}_2)/(m_1+m_2)$ is
the relative momentum between quark and antiquark.
$\mathbf{S}_A=\mathbf{s}_{q_1}+\mathbf{s}_{q_2}$,
$\mathbf{J}_A=\mathbf{L}_A+\mathbf{S}_A$ stand for the total spin
and total angular momentum, respectively. $\mathbf{L}_A$ is the
relative orbital angular momentum between $q_1$ and $\bar{q}_2$.
$\langle L_A M_{L_A} S_A M_{S_A}|J_A M_{J_A}\rangle$ denotes a
Clebsch-Gordan coefficient, and  $\chi^{12}_{S_AM_{S_A}}$,
$\phi^{12}_A$ and $\omega^{12}_A$ are the spin, flavor and color
wave functions, respectively.

The $S$-matrix of the process $A\rightarrow B + C$ is defined by
\begin{eqnarray}
\langle BC|S|A\rangle=I-2\pi i\delta(E_A-E_B-E_C)\langle
BC|T|A\rangle,
\end{eqnarray}
with
\begin{eqnarray}
\langle
BC|T|A\rangle=\delta^3(\mathbf{P}_A-\mathbf{P}_B-\mathbf{P}_C){\cal{M}}^{M_{J_A}M_{J_B}M_{J_C}},
\end{eqnarray}
where ${\cal{M}}^{M_{J_A}M_{J_B}M_{J_C}}$ is the helicity amplitude
of $A\rightarrow B + C$. In the center of mass frame of meson $A$,
$\mathbf{P}_A=0$, and  ${\cal{M}}^{M_{J_A}M_{J_B}M_{J_C}}$ can be
written as
\begin{widetext}
\begin{eqnarray}
{\cal{M}}^{M_{J_A}M_{J_B}M_{J_C}}(\mathbf{P})&=&\gamma\sqrt{8E_AE_BE_C}
\sum_{\renewcommand{\arraystretch}{.5}\begin{array}[t]{l}
\scriptstyle M_{L_A},M_{S_A},\\\scriptstyle M_{L_B},M_{S_B},\\
\scriptstyle M_{L_C},M_{S_C},m
\end{array}}\renewcommand{\arraystretch}{1}\!\!
\langle L_AM_{L_A}S_AM_{S_A}|J_AM_{J_A}\rangle
\langle L_BM_{L_B}S_BM_{S_B}|J_BM_{J_B}\rangle\nonumber\\
&&\times\langle
L_CM_{L_C}S_CM_{S_C}|J_CM_{J_C}\rangle\langle 1m1-m|00\rangle
\langle\chi^{14}_{S_BM_{S_B}}\chi^{32}_{S_CM_{S_C}}|\chi^{12}_{S_AM_{S_A}}\chi^{34}_{1-m}\rangle
\nonumber\\
&&\times[\langle
\phi^{14}_B\phi^{32}_C|\phi^{12}_A\phi^{34}_0\rangle \mathcal{I}^{M_{L_A},m}_{M_{L_B},M_{L_C}}(\mathbf{P},m_1,m_2,m_3)\nonumber\\
&&+(-1)^{1+S_A+S_B+S_C}\langle\phi^{32}_B\phi^{14}_C|\phi^{12}_A\phi^{34}_0\rangle
\mathcal{I}^{M_{L_A},m}_{M_{L_B},M_{L_C}}(-\mathbf{P},m_2,m_1,m_3)],
\end{eqnarray}
with the momentum space integral,
\begin{eqnarray}
\mathcal{I}^{M_{L_A},m}_{M_{L_B},M_{L_C}}(\mathbf{P},m_1,m_2,m_3)=\int
d\mathbf{p}\,\mbox{}\psi^\ast_{n_BL_BM_{L_B}}
({\scriptstyle\frac{m_3}{m_1+m_3}}\mathbf{P}+\mathbf{p})\psi^\ast_{n_CL_CM_{L_C}}
({\scriptstyle\frac{m_3}{m_2+m_3}}\mathbf{P}+\mathbf{p})
\psi_{n_AL_AM_{L_A}}
(\mathbf{P}+\mathbf{p}){\cal{Y}}^m_1(\mathbf{p}), \label{space}
\end{eqnarray}
\end{widetext} 
where $\mathbf{P}=\mathbf{P}_B=-\mathbf{P}_C$,
$\mathbf{p}=\mathbf{p}_3$, $m_3$ is the mass of the created quark
$q_3$; $\langle
\chi^{14}_{S_BM_{S_B}}\chi^{32}_{S_CM_{S_C}}|\chi^{12}_{S_AM_{S_A}}\chi^{34}_{1-m}\rangle$
and $\langle
\phi^{14}_B\phi^{32}_C|\phi^{12}_A\phi^{34}_0\rangle$  
are the overlap of spin and flavor wave function, respectively.

The spin overlap in terms of Winger's $9j$ symbol can be given by
\begin{eqnarray}
&&\langle
\chi^{14}_{S_BM_{S_B}}\chi^{32}_{S_CM_{S_C}}|\chi^{12}_{S_AM_{S_A}}\chi^{34}_{1-m}\rangle=\nonumber\\
&&\sum_{S,M_S}\langle S_BM_{S_B}S_CM_{S_C}|SM_S\rangle\langle
S_AM_{S_A}1-m|SM_S\rangle\nonumber\\
&&\times(-1)^{S_C+1}\sqrt{3(2S_A+1)(2S_B+1)(2S_C+1)}\nonumber\\
&&\times\left\{\begin{array}{ccc}
\frac{1}{2}&\frac{1}{2}&S_A\\
\frac{1}{2}&\frac{1}{2}&1\\
S_B&S_C&S
\end{array}\right\}.
\end{eqnarray}

Generally, one takes the simple harmonic oscillator (SHO)
approximation for the meson space wave functions in Eq.
(\ref{space}). In momentum-space, the SHO wave function reads
\begin{eqnarray}
&&\Psi_{nLM_L}(\mathbf{p})=(-1)^n(-i)^LR^{L+\frac{3}{2}}
\sqrt{\frac{2n!}{\Gamma(n+L+\frac{3}{2})}}\nonumber\\
&&~~~~\times\exp\left(-\frac{R^2p^2}{2}\right)L^{L+\frac{1}{2}}_n\left(R^2p^2\right)\mathcal{Y}_{LM_L}(\mathbf{p}),
\label{showave}
\end{eqnarray}
with
$\mathcal{Y}_{LM_L}(\mathbf{p})=|\mathbf{p}|^LY_{LM_L}(\Omega_p) $.
Here $R$ denotes the SHO wave function scale parameter; $\mathbf{p}$
represents the relative momentum between the quark and the antiquark
within a meson; $L^{L+\frac{1}{2}}_n\left(R^2p^2\right)$ is an
associated Laguerre polynomial.

The decay width for the process $A\to B + C$ in terms of the
helicity amplitude is
\begin{eqnarray*}
\Gamma=\pi^2\frac{|\mathbf{P}|^2}{M_A^2}\frac{1}{2J_A+1}
\sum_{\renewcommand{\arraystretch}{.5}\begin{array}[t]{l}
\scriptstyle M_{J_{M_A}},M_{J_{M_B}},\\
\scriptstyle \quad M_{J_{M_C}}
\end{array}}
\Big|\mathcal{M}^{M_{J_A}M_{J_B}M_{J_C}}\Big|^2\,.
\end{eqnarray*}

For comparing with experiments,
${\cal{M}}^{M_{J_A}M_{J_B}M_{J_C}}(\mathbf{P})$ can be converted
into the partial amplitude via the Jacob-Wick formula \cite{Jacob}
\begin{eqnarray}
&&{\mathcal{M}}^{J L}(A\rightarrow BC) = \frac{\sqrt{2 L+1}}{2 J_A
+1} \!\! \sum_{M_{J_B},M_{J_C}} \langle L 0 J M_{J_A}|J_A
M_{J_A}\rangle \nonumber\\&&\quad\quad\quad\times\langle J_B M_{J_B}
J_C M_{J_C} | J M_{J_A} \rangle \mathcal{M}^{M_{J_A} M_{J_B}
M_{J_C}}({\textbf{P}}),
\end{eqnarray}
where $\mathbf{J}=\mathbf{J}_B+\mathbf{J}_C$, $\mathbf{J}_{A}
=\mathbf{J}_{B}+\mathbf{J}_C+\mathbf{L}$ ,and $M_{J_A}=M_{J_B}+
M_{J_C}$. Then the decay width in terms of the partial wave
amplitude is taken as,
\begin{eqnarray}
\Gamma = \pi^2 \frac{{|\textbf{P}|}}{M_A^2}\sum_{JL}\Big
|\mathcal{M}^{J L}\Big|^2,\label{partialwidth}
\end{eqnarray}
where $|\textbf{P}|$, as mentioned above, is the three momentum of
the outgoing meson in the rest frame of meson $A$. According to the
calculation of 2-body phase space, one can get
\[|\textbf{P}|=\frac{\sqrt{[M^2_A-(M_B+M_C)^2][M^2_A-(M_B-M_C)^2]}}{2M_A},\]
where $M_A$, $M_B$, and $M_C$ are the masses of the meson $A$, $B$,
and $C$, respectively.

\section{The possible strong decay channels and amplitudes of the candidates for the $X(4160)$ and $X(3915)$\label{secamp}}

As analyzed in section \ref{introduction}, we consider the
$\eta_c(4^1S_0)$, $\chi_0(3^3P_0)$, $\chi_1(3^3P_1)$, $\eta_{c2}(2^1D_2) $
as the possible candidates of the $X(4160)$, and assume that the
upper limit of the mass is 4156 MeV observed by Belle.  For the
X(3915), one chooses charmonium state $\chi_0(2^3P_0)$ with mass
3916 MeV. According to the $^3P_0$ model discussed in the above section,
the OZI rule allows open-charm strong decay and corresponding
amplitudes of possible charmonium states are listed in Tables
\ref{decay mode} and \ref{amplitude}. We replace
$\mathcal{I}^{+1-1}_{0,0}, \mathcal{I}^{-1+1}_{0,0}$ with
$\mathcal{I}^{\pm}$ and $\mathcal{I}^{0,0}_{0,0}$ with
$\mathcal{I}^{0,0}$ in Table \ref{amplitude}, respectively. The
details of the spatial integral about
$\mathcal{I}^{\pm}(\mathbf{P})$ and $\mathcal{I}^{0,0}(\mathbf{P})$
are given in the Appendix.
\begin{ruledtabular}
\begin{table}[htb]
\caption{The OZI rule and phase space allowed open-charm strong
decay modes of the possible charmonium states for the X(4160) and
X(3915).  \label{decay
mode}}
\begin{tabular}{c|c|c|lcc}
  State     &$J^{PC}$            &Decay mode    &Decay channel\\
\hline
  $\eta_c(4^1S_0)$ &$0^{-+}$  &$0^- + 1^-$    &
  $D\bar{D}^*,D_s^+D_s^{*-}$\\
                  & &$1^- + 1^-$    & $D^*\bar{D}^*$
  \\\hline
  $\chi_0(3^3P_0)$       &$0^{++}$  &$0^- + 0^-$    &
  $D\bar{D},D_s^+D_s^{-}$\\
                  & &$1^- + 1^-$    & $D^*\bar{D}^*$\\\hline
$\chi_1(3^3P_1)$       &$1^{++}$  &$0^- + 1^-$    &
  $D\bar{D}^*,D_s^+D_s^{*-}$\\
                  & &$1^- + 1^-$    & $D^*\bar{D}^*$
                  \\\hline
$\eta_{c2}(2^1D_2)$       &$2^{-+}$  &$0^- + 1^-$    &
  $D\bar{D}^*,D_s^+D_s^{*-}$\\
                  & &$1^- + 1^-$    & $D^*\bar{D}^*$
                   \\\hline
 $ \chi_0(2^3P_0)$       &$0^{++}$  &$0^- + 0^-$    &
  $D\bar{D}$
\end{tabular}
\end{table}
\end{ruledtabular}

\begin{ruledtabular}
\begin{table}[htb]
\caption{The partial wave amplitude for the strong decays of
relevant charmonium state. The element of flavor matrix $\langle
\phi^{14}_B\phi^{32}_C|\phi^{12}_A\phi^{34}_0\rangle=1/\sqrt{3}$ in
present work. We take $\mathcal{E}=\gamma\sqrt{E_AE_BE_C}$ in this
table. \label{amplitude}}
\begin{tabular}{c|cc}
  State       &decay channel            &Decay amplitude\\
\hline
  $\eta_c(4^1S_0)$ &$0^- + 1^-$   & $\mathcal{M}^{11}=\frac{\sqrt{2}}{3} \mathcal{E}\mathcal{I}^{00}$
  \\
                   &$1^- + 1^-$   & $\mathcal{M}^{11}=\frac{2}{3} \mathcal{E}\mathcal{I}^{00}$ \\\hline
  $\chi_0(3^3P_0)$ &$0^- + 0^-$   & $\mathcal{M}^{00}=\frac{\sqrt{2}}{3\sqrt{3}} \mathcal{E}\left(\mathcal{I}^{00}-2\mathcal{I}^{\pm}\right)$  \\
                   &$1^- + 1^-$   & $\mathcal{M}^{00}=\frac{\sqrt{2}}{9} \mathcal{E}\left(\mathcal{I}^{00}-2\mathcal{I}^{\pm}\right)$\\
                    &             & $\mathcal{M}^{22}=\frac{4}{9} \mathcal{E}\left(\mathcal{I}^{00}+\mathcal{I}^{\pm}\right)$\\
                   \hline
 $\chi_1(3^3P_1)$  &$0^- + 1^-$   & $\mathcal{M}^{10}=\frac{2}{9} \mathcal{E}\left(\mathcal{I}^{00}-2\mathcal{I}^{\pm}\right)$\\
                   &              & $\mathcal{M}^{12}=\frac{\sqrt{2}}{9} \mathcal{E}\left(\mathcal{I}^{00}+\mathcal{I}^{\pm}\right)$\\
                   &$1^- + 1^-$   & $\mathcal{M}^{22}=\frac{2}{3\sqrt{3}}
                   \mathcal{E}\left(\mathcal{I}^{00}+\mathcal{I}^{\pm}\right)$\\\hline
$\eta_{c2}(2^1D_2)$&$0^- + 1^-$   & $\mathcal{M}^{11}=\frac{2}{15} \mathcal{E}\left(\sqrt{3}\mathcal{I}^{\pm}-\mathcal{I}^{00}\right)$\\
                   &$1^- + 1^-$   & $\mathcal{M}^{11}=\frac{2\sqrt{2}}{15}
                   \mathcal{E}\left(\sqrt{3}\mathcal{I}^{\pm}-\mathcal{I}^{00}\right)$\\\hline
$\chi_0(2^3P_0)$ &$0^- + 0^-$   & $\mathcal{M}^{00}=\frac{\sqrt{2}}{3\sqrt{3}} \mathcal{E}\left(\mathcal{I}^{00}-2\mathcal{I}^{\pm}\right)$  \\
\end{tabular}
\end{table}
\end{ruledtabular}

\section{Numerical results and discussion \label{discussion}}
There are several parameters should be input to calculate the strong
decay in the $^3P_0$ model. In the present work, the masses
of constituent quarks are taken as $m_{u}=m_d=0.22$ GeV, $m_s=0.419$ GeV,
$m_c=1.6$ GeV \cite{Godfrey}. The strength of quark pair creation
$\gamma=6.95$ has been adopted by many literatures
\cite{Blundell,zhusl}, which is fitted by strong decay of
light-, charmonium-, open charmed-mesons and baryons observed by
experiments. The value of $\gamma$ is higher than that used in Ref.
\cite{isgur2} by a factor of $\sqrt{96\pi}$ due to different field
theory conventions. The strength of $s\bar{s}$ creation satisfies
$\gamma_s=\gamma/\sqrt{3}$ \cite{Yaouanc2}. Refs.\cite{XiangLiu,zhusl,XiangLiu2}
also take this value to study the
strong decay of charmonium, heavy-light meson and heavy baryons. In
this work, we take these parameters for calculation as well. The $R$
values of $D,~D^*,~D_s,~D_s^*$ in the SHO are shown in Table
\ref{parameter}, which are obtained by the calculation of the nonrelativistic quark
model with Coulomb item, linear confinement and smeared hyperfine interactions.
\begin{ruledtabular}
  \begin{table}[htb]
   \caption{The parameters relevant to the two-body strong decays of the charmonium state in the $^3P_0$ model.
\label{parameter}}
    \begin{tabular}{cccccc}
    State &Mass (MeV) \cite{PDG}&$R$ (GeV$^{-1}$) \cite{Godfrey}\\\hline
    $D$&$1869.62(\pm)$~~1864.84(0)&1.52 \\
    $D^*$&$2021.27(\pm)$~~2006.97(0)&1.85 \\
    $D_s$&$1968.49(\pm)$&1.41\\
    $D_s^*$&$2112.3(\pm)$&1.69\\
    \end{tabular}
    \end{table}
\end{ruledtabular}

First of all, we study the strong decay of the $\chi_0(3^3P_0)$
which is discussed by Chao and Li in Refs.\cite{chaox4160,prd79094004}
from the production process of $e^+e^-
\to J/\psi + X(4160)$ and the mass spectrum is obtained by
the potential model with color screening. Using the method of
Numerov algorithm \cite{Koonin}, we also obtain the mass
4149 MeV by the same potential and parameters in Ref.
\cite{prd79094004}. Usually, the width of strong decay is sensitive
\cite{Close,XiangLiu,Swanson,zhusl,Blundell,ldm} to the $R$ value in
the SHO. Here the reasonable value of $R$ is obtained by fitting
the wave function obtained by solving the schr\"{o}dinger equation
\cite{prd79094004}.

Through the Fourier transform, the Eq. (\ref{showave}) turns into
\begin{eqnarray}
\Psi_{nLM_L}(\mathbf{r})=R_{nL}(r)Y_{LM_L}(\Omega_r), \label{sho}
\end{eqnarray}
with the radial wave function
\begin{eqnarray}
&&R_{nL}(r)=R^{-(L+\frac{3}{2})}
\sqrt{\frac{2n!}{\Gamma(n+L+\frac{3}{2})}}\nonumber\\
&&~~~~~\times\exp\left(-\frac{R^{-2}r^2}{2}\right)r^LL^{L+\frac{1}{2}}_n\left(R^{-2}r^2\right).
\label{Rnl}
\end{eqnarray}
The wave function $u(r)=r~R_{nL}(r)$ of charmonium state $3P$ is
shown in Fig.2. Using Eq.(\ref{Rnl}) to fit the wave
function got by Numerov algorithm method (the wave function is denoted as 'NAWF'
in the following), we can get the $R=2.5\sim2.98$ GeV$^{-1}$.
\begin{center}
\epsfig{file=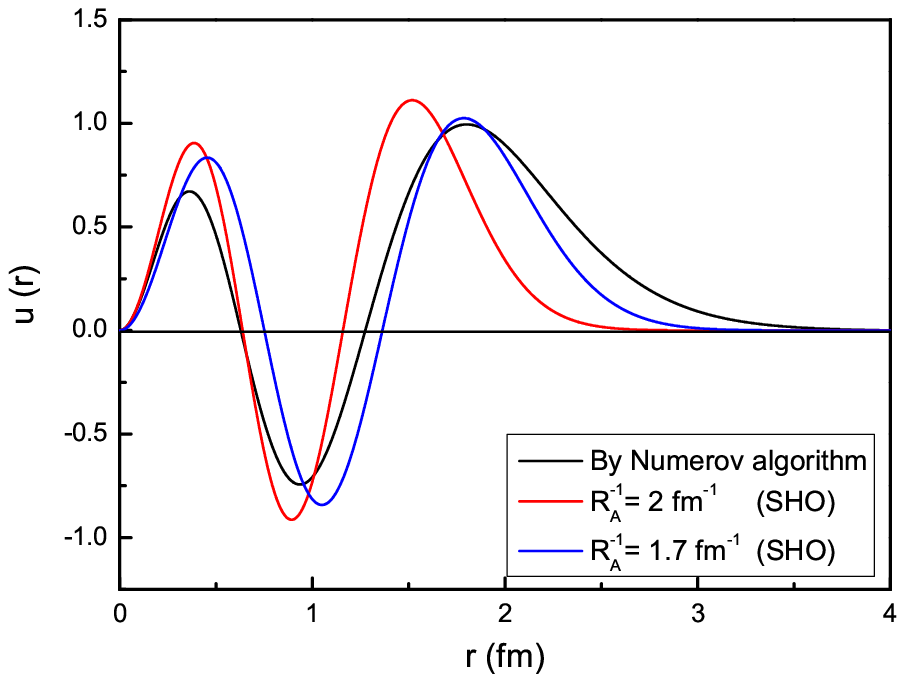,width=8.6cm}

{\small Fig.2 The wave function of charmonium state $3P$. }
\end{center}

The $\chi_0(3^3P_0)$ has decay channels of $0^{++} \to 0^-+0^-$ with
$S$-wave and  $0^{++} \to 1^-+1^-$ with $S$-,$~D$-wave, while the
$0^{++} \to 0^-+1^-$ is forbidden. Therefore, it can decay into
$D\bar{D},~D_sD_s,~D^*\bar{D}^*$, which are allowed by the phase
space. In Fig.3, we show the dependence of the
partial widths of the strong decay of the $\chi_0(3^3P_0)$ on the
$R_A$. Taking $R_A=2.5\sim2.98$ GeV$^{-1}$ discussed above, the
total width ranges from 105 to 143 MeV which falls in the range of
experimental data. However, the dominate contribution comes from the
$\chi_0(3^3P_0)\to DD$ which is inconsistent with the experimental
result. So the assignment of the charmonium state $\chi_0(3^3P_0)$ to
the $X(4160)$ is disfavored.
\begin{center}
\epsfig{file=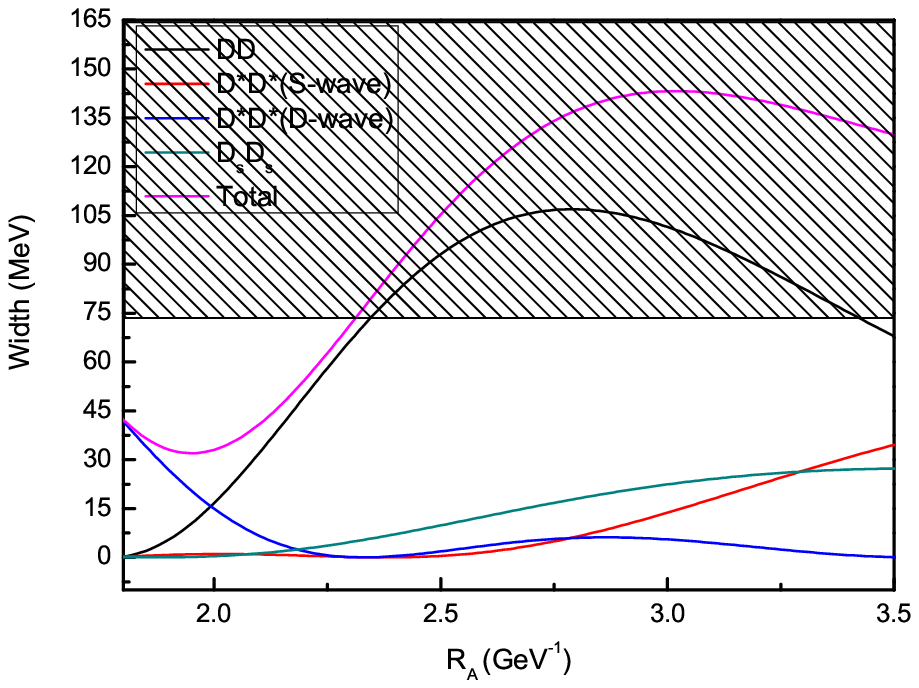,width=8.6cm}
{\small Fig.3 The possible strong decay of the $\chi_0(3^3P_0)$.}
\end{center}

The $\eta_c(4^1S_0)$ is mostly like the $X(4160)$ for it has high
production cross sections in the process of $e^+e^- \to J/\psi +
X(4160)$ discussed by Chao \cite{chaox4160}. However, it is
difficult to understand why the predicted mass 4250 MeV
\cite{prd79094004}, 4384, 4425 MeV \cite{prd72054026} are much
higher than 4156 MeV. By considering the effect of the meson loops
\cite{our}, the mass may be lower than that of
Refs.\cite{prd79094004,prd72054026}. Here, we assume the mass of the
$\eta_c(4^1S_0)$ is 4156 MeV. The main decay channels of the
$\eta_c(4^1S_0)$ are $0^{-+}\to 0^-+1^-$ and $0^{-+}\to 1^-+1^-$
with $P$-wave between outgoing mesons. Obviously, the $0^{-+}\to
0^-+0^-$ is forbidden. The decay width of main decay channels are
shown in Fig.4. The total width can only reach up to
about 25 MeV with $R_A$ around 2.9 GeV, which is obtained by fitting to
NAWF of the $\eta_c(4^1S_0)$. It is about 3 times smaller than the
lower limit of the experimental result of the $X(4160)$. Since the results of
some hadron states predicted by the $^3P_0$ model may be a factor of
$2 \sim 3$ off the experimental width due to inherent uncertainties
of this model \cite{NPB10521,Yaouanc,PRD546811,Roberts1,Blundell}, the assignment
of the X(4160) to the $\eta_c(4^1S_0)$ cannot be excluded.
The ratio of main decay channel $D\bar{D}^*,~D^*\bar{D}^*$ is
\begin{eqnarray}
\frac{\mathcal{B}(\eta_c(4^1S_0) \to
D\bar{D}^*)}{\mathcal{B}(\eta_c(4^1S_0) \to D^*\bar{D}^*)} =1.25 .
\end{eqnarray}
It is much larger than the $0.22$ reported by Belle. If one takes
the $\eta_c(4^1S_0)$ as an assignment of X(4160), the precision
measurement of  the ratio between the width of the $D\bar{D}^*$ and
$D^*\bar{D}^*$ is necessary in further experiment.
\begin{center}
\epsfig{file=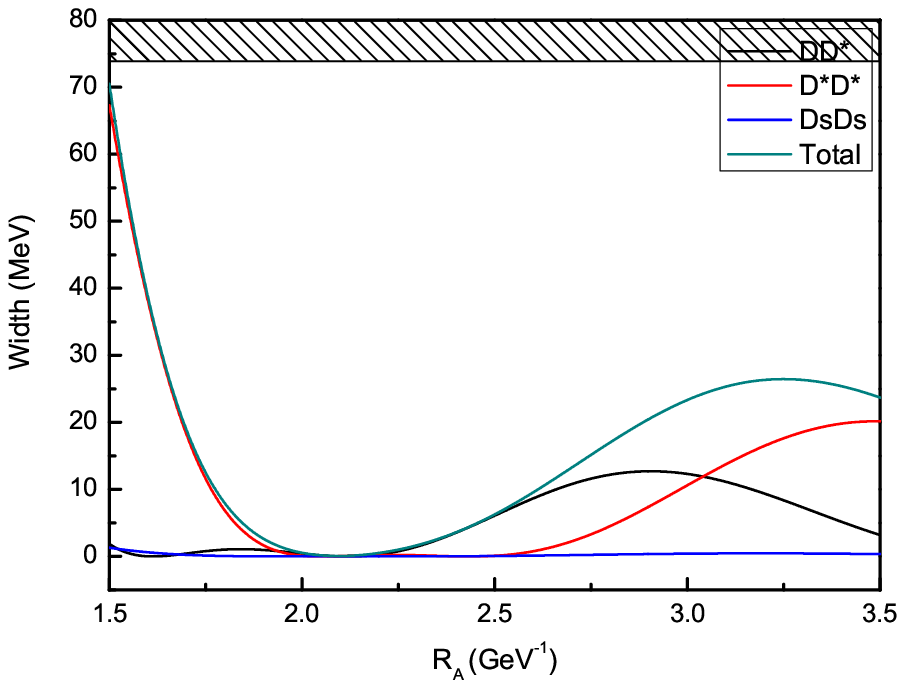,width=8.6cm}

{\small Fig.4 The possible strong decay of the $\eta_c(4^1S_0)$.}
\end{center}

Because the $\chi_1(3^3P_1)$ has quantum number $J^{PC}=1^{++}$ and
mass 4178 MeV, it is also a possible candidate of the $X(4160)$.
$1^{++}\to 0^-+1^-$ and $1^{++}\to 1^-+1^-$ with $S$- and $D$-wave
are the main decay channels of the $\chi_1(3^3P_1)$. Fig.5
shows our results in the $^3P_0$ model. Taking
$R_A=2.5\sim2.98$ GeV$^{-1}$, the total width is consistent with the
range of the $X(4160)$. However, the dominant decay is
$\chi_1(3^3P_1) \to D\bar{D}^*$ while the decay width has only a few
MeV for the $\chi_1(3^3P_1) \to D^*\bar{D}^*$ channel, which is
inconsistent with the experimental data. Therefore, regarding the
$X(4160)$ as the $\chi_1(3^3P_1)$ state is impossible.
\begin{center}
\epsfig{file=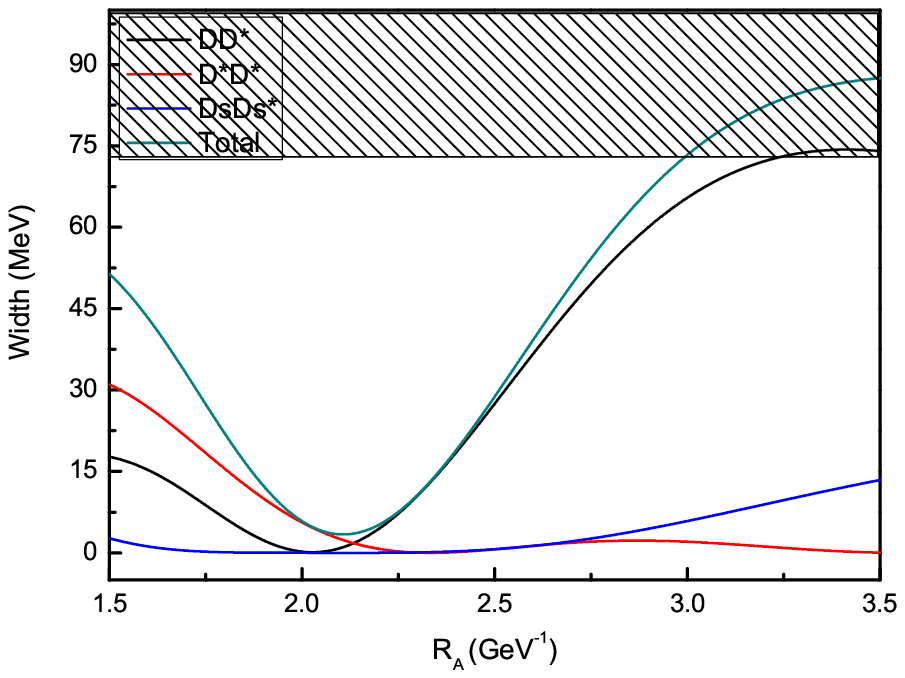,width=8.6cm}

{\small Fig.5 The possible strong decay of the $\chi_1(3^3P_1)$.}
\end{center}

The another possible candidate of the $X(4160)$ is the charmonium
state $\eta_{c2}(2^1D_2)$. Firstly, it has quantum number
$J^{PC}=2^{-+}$ and mass 4099 MeV \cite{prd79094004}, 4158 MeV \cite
{prd72054026} which are compatible with the result of Belle.
Secondly, the $\psi(4160)$\cite{PDG} is known to be the good
candidate of the $\psi(2^3D_1)$ with $J^{PC}=1^{--}$, which is
discussed in detail by Chao \cite{chaox4160}. So the $X(4160)$ may
be the D-wave spin-singlet charmonium state $^1D_2(2D)$. Thirdly,
$\eta_{c2}(2^1D_2)$ decaying into $D\bar{D}$ is forbidden, and this
decay is also not seen by Belle.

For the strong decay of the $\eta_{c2}(2^1D_2)$, it has $2^{-+}\to
0^-+1^-$ and $2^{-+}\to 1^-+1^-$ decay channels with $P$-wave
between outgoing mesons. In this case, final states
$D\bar{D}^*,~D_s\bar{D}_s^*$ and $D^*\bar{D}^*$ are phase space
allowed. In Fig.6, we present the numerical results
of main decay channels for the $\eta_{c2}(2^1D_2)$. By fitting the
NAWF of the $\eta_{c2}(2^1D_2)$, we get $R_A=2.7 \sim 3.0 $
GeV$^{-1}$. The total decay width of the $\eta_{c2}(2^1D_2)$ falls
in the range of the $X(4160)$ released by Belle. Taking the
reasonable $R_A$ value of the SHO, the ratio of the main decay
channel
 $D\bar{D}^*,~D^*\bar{D}^*$ is
\begin{eqnarray}
\frac{\mathcal{B}(\eta_{c2}(2^1D_2) \to
D\bar{D}^*)}{\mathcal{B}(\eta_{c2}(2^1D_2)) \to D^*\bar{D}^*)} =1.4
\sim 0.76
\end{eqnarray}
and shown in Fig.7. However, the result is
somewhat larger than the
$\mathcal{B}_{D^*\bar{D}}(X(4160))/\mathcal{B}_{D^*\bar{D}^*}(X(4160))<
0.22$ observed by Belle. We believe that to measure this ratio is
very important since it is independent on the uncertain strength
$\gamma$ of the quark pair creation from vacuum.

To sum up, the $\eta_{c2}(2^1D_2)$ is a better candidate for the
$X(4160)$ in the present calculation.
\begin{center}
\epsfig{file=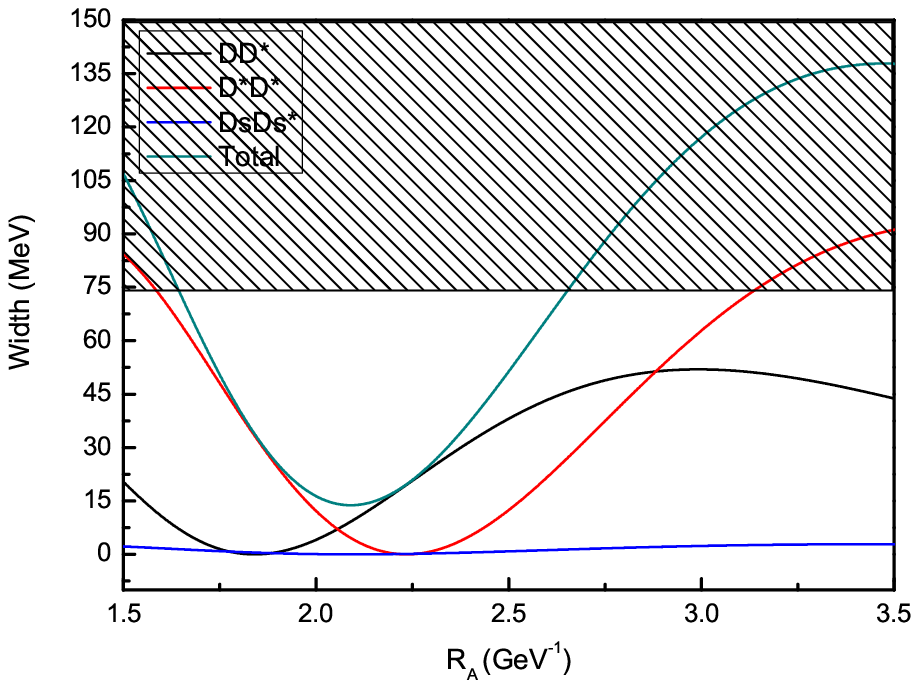,width=8.6cm}

{\small Fig.6 The possible strong decay of the $\eta_{c2}(2^1D_2)$.}
\end{center}

\begin{center}
\epsfig{file=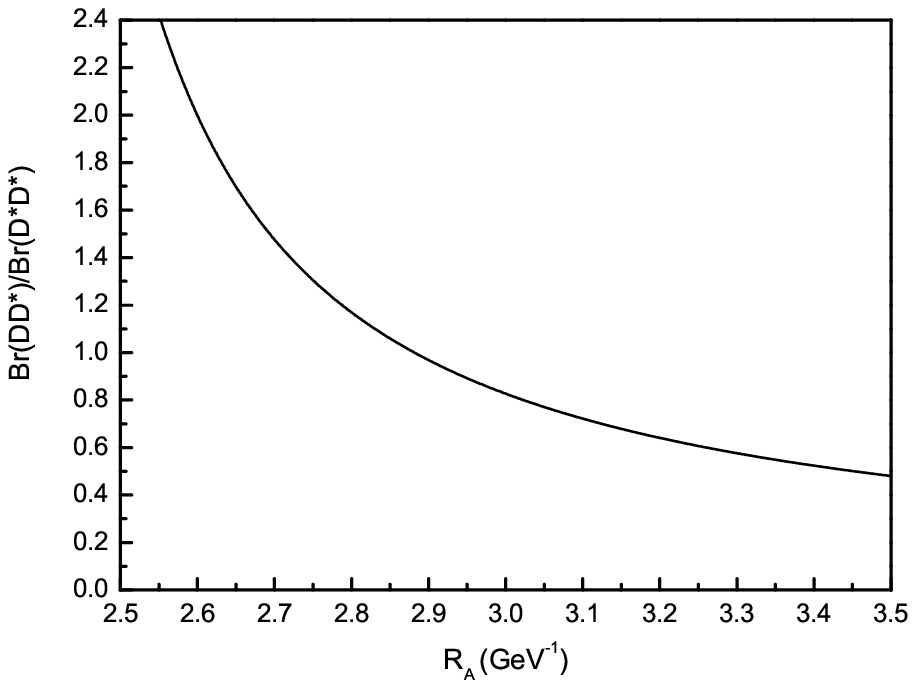,width=8.6cm}

{\small Fig.7 The ration of $\frac{\mathcal{B}(\eta_{c2}(2^1D_2) \to
D\bar{D}^*)}{\mathcal{B}(\eta_{c2}(2^1D_2)) \to D^*\bar{D}^*)}$ with
$R_A$ value of the SHO.}
\end{center}

The $X(3915)$, which was observed by Belle in $\gamma\gamma \to
\omega J/\psi$ with a statical significance of $7.5\sigma$
\cite{Olsen0909}, is the most recent addition to the collection of
the $XYZ$ states. According to the Table \ref{ccspectrum} predicted
by potential model, the excited charmonium state $\chi_0(2^3P_0)$ is
a good candidate for the $X(3915)$, due to it has mass
$M=3914\pm4\pm2$ MeV and the possible quantum number is
$J^{PC}=0^{++}$.

The $\chi_0(2^3P_0)$ has only the strong decay channel $0^{++}\to
0^-+0^-$ allowed by phase space. The width of $\chi_0(2^3P_0) \to
D\bar{D}$ with $R_A$ of the SHO is presented in Fig.8.
The total width ranges from 132 to 187 MeV with
$R_A=2.3 \sim 2.5$ GeV$^{-1}$ fitted to the NAWF of the
$\chi_0(2^3P_0)$. It is much larger than the $\Gamma
=28\pm12^{+2}_{-8}$ MeV reported by Refs.
\cite{Olsen0909,CZYuan0910,Zupanc0910}. Therefore, the $X(3915)$ is
unlikely to be the charmonium state $\chi_0(2^3P_0)$ although the
mass is compatible with the $X(3915)$.
\begin{center}
\epsfig{file=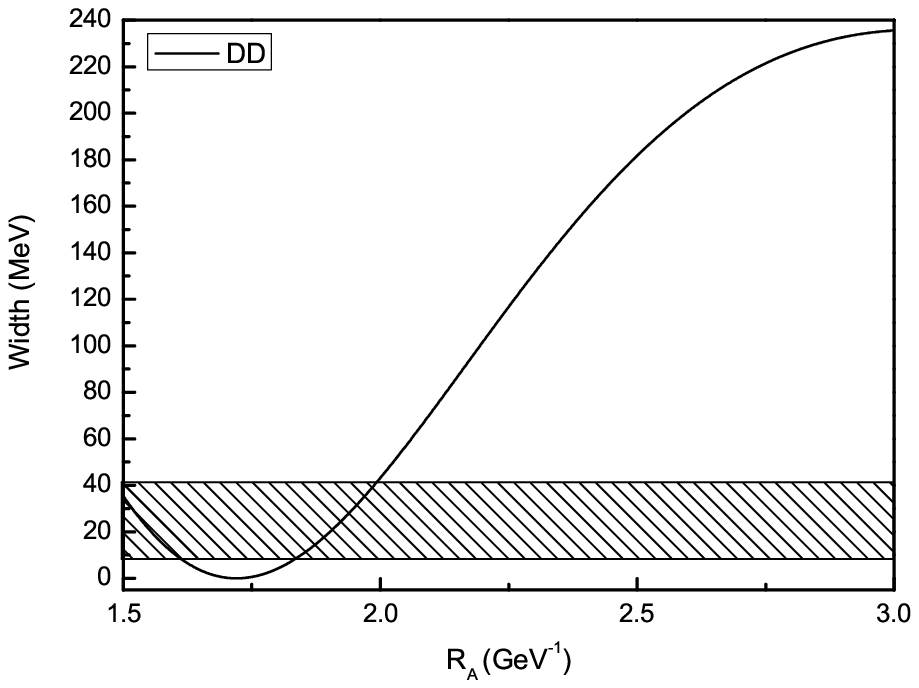,width=8.6cm}

{\small Fig.8 The possible strong decay of the $\chi_0(2^3P_0)$.}
\end{center}

\section{Summery \label{summery}}
In summary, we have discussed the possible interpretations of the
$X(4160)$ observed by Belle collaborations in $e^+e^- \to J/\psi + X(4160)$
followed by $X(4160)\to D^*\bar{D}^*$. We also study the newest
state $X(3915)$ observed by Belle in the process $\gamma\gamma \to
J/\psi \omega$ \cite{Olsen0909}.

In quark models, the masses of the charmonium states:
$\chi_0(3^3P_0)$, $\chi_1(3^3P_1)$, $\eta_{c2}(2^1D_2)$ are all around
4156 MeV. By taking the effect of virtual mesons loop \cite{our}
into account, the $\eta_c(4^1S_0)$ may also has mass around 4156
MeV. All the four states have charge parity $C=+$ which are
compatible with the $X(4160)$ observed by Belle.

For the strong decay of the $\chi_0(3^3P_0)$, the dominant strong
decay is $\chi_0(3^3P_0) \to D\bar{D}$ while $\chi_0(3^3P_0) \to
D^*\bar{D}^*$ contributes to the total width only a little in the
reasonable $R$ in the SHO. It is contrast to the experimental
result. Thus the excited charmonium state $\chi_0(3^3P_0)$ disfavor
the $X(4160)$.

The $\eta_c(4^1S_0)$ can not decay into $D\bar{D}$ and may has high
production rate \cite{chaox4160} in $e^+e^- \to J/\psi +\eta_c(4S)$
process by analogy with $e^+e^- \to J/\psi +\eta_c(1S)(\eta_c(2S)
\chi_{c0}(1P))$. However, the total width in present work is lower
than the experimental data of the $X(4160)$.

The main strong decay channel of the $\chi_1(3^3P_1)$ is
$D\bar{D}^*$ while $D^*\bar{D}^*$ is only a few MeV. It is
inconsistent with the results of Belle. Therefore, taking the
$\chi_1(3^3P_1)$ as an assignment for the $X(4160)$ is impossible.

The $\eta_{c2}(2^1D_2)$ can not decay to $D\bar{D}$ which is also
not seen in the experiment. The total width of the
$\eta_{c2}(2^1D_2)$ match well with the data of the $X(4160)$ in our
calculation.
So, the $\eta_{c2}(2^1D_2)$ is a good candidate for the
$X(4160)$, for it is not only the mass but also the strong decay are
well compatible with the results observed by Belle, although the excited
charmonium state $\eta_c(4^1S_0)$ can not be rule out as an
assignment for the $X(4160)$.

We also give the ratio of
$\frac{\mathcal{B}(\eta_{c2}(2^1D_2) \to
D\bar{D}^*)}{\mathcal{B}(\eta_{c2}(2^1D_2)) \to D^*\bar{D}^*)}$
which is independent on the parameter $\gamma$ in the $^3P_0$ model.
The numerical result is somewhat larger than the experimental data.
Therefore, we suggest Belle, BaBar and other experimental
collaborations to measure it to confirm this state.

By assuming the $X(3915)$ is the $\chi_0(2^3P_0)$, the
strong decay of the state is calculated. From our numerical results, we think this assumption
is unacceptable. Due to the partial width of the $X(3915)$ to
$\gamma\gamma$ or $\omega J/\psi$ is too large, Yuan
\cite{CZYuan0910} also believes that it is very unlikely to be a
charmonium state. Thus, It is necessary to do more study
to understand the properties of the $X(3915)$.

\acknowledgments{
 You-chang Yang would like to thank Xin Liu for
useful discussion. The work is supported partly by the National
Science Foundation of China under Contract No.10775072 and the
Research Fund for the Doctoral Program of Higher Education of China
under Grant No. 20070319007, No. 1243211601028.}

\begin{widetext}
\section*{Appendix}
The spatial overlap
$\mathcal{I}_{M_{L_B},M_{L_C}}^{M_{L_A},m}(\textbf{P},m_1,m_2,m_3)$
is simplified as $\mathcal{I}^{n'm'}(\textbf{P})$ in present work
due to $M_{L_B}=M_{L_C}=0$. According to the Eq. (\ref{space}), the
concrete calculations of the integration are trivial after choosing
the direction of $\textbf{P}$ along $z$ axis \cite{Jacob}.  We list
all expressions of $I^{\pm},I^{00}$ used in Table \ref{amplitude}

In the case of $2P\to 1S + 1S$
\begin{eqnarray}
&&I^{\pm}=I^{1-1}=I^{-11}\nonumber\\
&&\quad=i\frac{\sqrt{6}}{\sqrt{5}\pi^{5/4}\Delta^7}\left(R_A^{5/2}R_B^{3/2}R_C^{3/2}\right)
\exp\left(-\frac{1}{2}\zeta^2\textbf{P}^2\right)(10~ R_A^2+\Delta^2(-5+2~ \textbf{P}^2 R_A^2 (1+\lambda)^2))\nonumber\\
&&I^{00}=-i\frac{\sqrt{6}}{\sqrt{5}\pi^{5/4}\Delta^7}
\left(R_A^{5/2}R_B^{3/2}R_C^{3/2}\right)\exp\left(-\frac{1}{2}\zeta^2\textbf{P}^2\right)\nonumber\\
&&~~~~(10R_A^2+\Delta^2(-5+\textbf{P}^2(1+\lambda)(-5~ \Delta^2
\lambda + 2~ R_A^2 (3+\lambda(8+\Delta^2
\textbf{P}^2(1+\lambda)^2))))). \label{am2Pto1s1s}
\end{eqnarray}

For $2D\to 1S + 1S$
\begin{eqnarray}
&&I^{\pm}=I^{1-1}=I^{-11}\nonumber\\
&&\quad=\frac{2 \sqrt{3}}{\sqrt{7}\pi^{5/4}\Delta^7}\left(R_A^{7/2}R_B^{3/2}R_C^{3/2}\right)
\exp\left(-\frac{1}{2}\zeta^2\textbf{P}^2\right)\textbf{P}(1+\lambda)(14~R_A^2+\Delta^2(-7+2~\textbf{P}^2 R_A^2 (1+\lambda)^2))\nonumber\\
&&I^{00}=-\frac{2}{\sqrt{7}\pi^{5/4}\Delta^7}
\left(R_A^{7/2}R_B^{3/2}R_C^{3/2}\right)\exp\left(-\frac{1}{2}\zeta^2\textbf{P}^2\right)\textbf{P}(1+\lambda)(28~R_A^2+\Delta^2(-14+\textbf{P}^2(1+\lambda)\nonumber\\
&&~~~~~~(-7~ \Delta^2 \lambda + 2~ R_A^2 (4+\lambda(11+\Delta^2
\textbf{P}^2(1+\lambda)^2))))). \label{am2Dto1s1s}
\end{eqnarray}

For $3P\to 1S + 1S$
\begin{eqnarray}
&&I^{\pm}=I^{1-1}=I^{-11}\nonumber\\
&&\quad=i\frac{\sqrt{3}}{\sqrt{70}\pi^{5/4}\Delta^9}\left(R_A^{5/2}R_B^{3/2}R_C^{3/2}\right)
\exp\left(-\frac{1}{2}\zeta^2\textbf{P}^2\right)(140~R_A^4+28~\Delta^2R_A^2(-5+2~\mathbf{P}^2R_A^2(1+\lambda)^2)\nonumber\\
&&~~~~~~~+\Delta^4(35-28~\mathbf{P}^2R_A^2(1+\lambda)^2+4~\mathbf{P}^4R_A^4(1+\lambda)^4))\nonumber\\
&&I^{00}=-i\frac{2\sqrt{6}}{\sqrt{35}\pi^{5/4}\Delta^9}
\left(R_A^{5/2}R_B^{3/2}R_C^{3/2}\right)\exp\left(-\frac{1}{2}\zeta^2\textbf{P}^2\right)(35~R_A^4+\frac{1}{4}~\Delta^6\mathbf{P}^2
\lambda(1+\lambda)\nonumber\\
&&~~~~~~(35-28~\mathbf{P}^2R_A^2(1+\lambda)^2+4~\mathbf{P}^4R_A^4(1+\lambda)^4)+7~\Delta^2R_A^2(-5+\mathbf{P}^2R_A^2(1+\lambda)(6+11~\lambda))\nonumber\\
&&~~~~~~+\frac{1}{4}~\Delta^4(35-28~\mathbf{P}^2R_A^2(1+\lambda)(3+8~\lambda)+4~\mathbf{P}^4R_A^4(1+\lambda)^3(5+19~\lambda))).
\label{am3Pto1s1s}
\end{eqnarray}

For $4S\to 1S + 1S$
\begin{eqnarray}
&&I^{00}=\frac{1}{2\sqrt{120}\pi^{5/4}\Delta^9}
\left(R_A^{3/2}R_B^{3/2}R_C^{3/2}\right)\exp\left(-\frac{1}{2}\zeta^2\textbf{P}^2\right)\textbf{P}
(840~R_A^6(2+3~\lambda)+\Delta^6\lambda(-105+210~\mathbf{P}^2R_A^2(1+\lambda)^2\nonumber\\
&&~~~-84~\mathbf{P}^4R_A^4(1+\lambda)^4+8~\mathbf{P}^6R_A^6(1+\lambda)^6)+6~\Delta^4R_A^2(70+175~\lambda-28~\mathbf{P}^2R_A^2(1+\lambda)^2(2+7~\lambda)\nonumber\\
&&~~~+4~\mathbf{P}^4R_A^4(1+\lambda)^4(2+9~\lambda))+84~\Delta^2
R_A^4(-5(4+7~\lambda)+2~\mathbf{P}^2R_A^2(1+\lambda)^2(4+9~\lambda))).\label{am4Sto1s1s}
\end{eqnarray}

Here, The parameters $\Delta$, $\zeta$ and $\eta$  in Eqs.
(\ref{am2Pto1s1s}), (\ref{am2Dto1s1s}), (\ref{am3Pto1s1s}),
(\ref{am4Sto1s1s}) are defined as
\begin{eqnarray*}
&&\Delta^2=R_A^2+R_B^2+R_C^2, \,\,
\lambda=-\frac{R_A^2+\xi_1R_B^2+\xi_2R_C^2}{R_A^2+R_B^2+R_C^2},\nonumber\\
&&\zeta^2=R_A^2+\xi_1^2R_B^2+\xi_2^2R_C^2-\frac{(R_A^2+\xi_1R_B^2+\xi_2R_C^2)^2}{R_A^2+R_B^2+R_C^2}.\nonumber\\
\end{eqnarray*}
with
\begin{eqnarray*}
\xi_1=\frac{m_3}{m_3+m_1}, \quad \xi_2=\frac{m_3}{m_3+m_2}.
\end{eqnarray*}
Here $m_1, m_2$ and $m_3$ denotes the mass of quark inside parent
meson and created from vacuum, respectively.
\end{widetext}


\begin{thebibliography}{99}
\bibitem{our}This work is in progress by our group.
\end{thebibliography}
\end{document}